# Analysis of access in the Take-Grant model

## 1. Introduction

One of the minimum requirement of computer system security is the presence of access control to information [1]. This means that the system should be organized so that users have access only to information they need to perform their duties. For this purpose developed a number of mathematical models for access control.

One of the most thoroughly-developed models is the Take-Grant protection model [2]. The central concept of this model is a ***safe state.*** The model identified criteria of safety states. This article offers several methods for determining the safety of the computer system state based on Take-Grant protection model.

## 2. Description of the Take-Grant model

A computer system in the Take-Grant model is represented as a directed graph (called a ***protection graph***) whose vertices are the subjects and objects of the computer system and arcs are the access right set of subjects to objects. Addition to the standard access rights such as the read or write access to the Take-Grant introduces two additional access rights: ***Take (t)*** is the right to take any access rights from the subject, and ***Grant (g)*** is the right to assign its access rights to any subject. An analysis of the original graph can determine is it possible to a particular subject to gain access to a particular object for a certain number of steps. It means the analysis allows us to determine the possibility to modify the original graph, so that between two vertices will be the arc which did not exist in the original graph.

There are two main concepts followed from the Take-Grant model that describes the conditions of access. The first concept is formulated for a graph containing only the subjects as vertices. The essence of this concept is that the access between the two vertices will be possible if there are the path in original graph between this vertices which running through the other vertices and contains access rights t or g (***tg-path***) [2]. The second concept is formulated for an arbitrary protection graph. In order to understand its essence we must give a number of additional definitions.

*Island* in an arbitrary protection graph is the tg-connected subject only subgraph.

*The bridge* is the tg-path running through objects, having a certain type: $\overrightarrow{t^*}$, $\overleftarrow{t^*}$, $\overrightarrow{t^*}\overrightarrow{g}\overleftarrow{t^*}$ or $\overrightarrow{t^*}\overleftarrow{g}\overleftarrow{t^*}$, here symbol "*" means multiple repeat (including zero).

*Initially* and *terminally spans of the bridge* are pathes running through objects which has the form of $\overrightarrow{t^*}\overrightarrow{g}$ and $\overrightarrow{t^*}$ respectively.

The essence of the second concept is that the access between the two vertices in the protection graph is possible in the case when those vertices are connected in the original graph with the islands by the initially and terminally span of the bridge, respectively, and the islands are connected with each other by a bridge.

## 3. Investigation of system states safety

The theoretical description of the Take-Grant model gives conditions of access. Further in the article describes ways to verify the conditions of access different from those in the classical model.

If the protection graph consists only of subject vertex it is necessary to check the presence of tg-path between two vertices. For this purpose, we use Dijkstra's algorithm [3] as follows:

1) According to the original protection graph $G_0$ construct a graph $G_0^*$ as follows:

   a) Set of vertices of the graph $G_0^*$ is the same as set of vertices of the graph $G_0$ ( $V_0^* \equiv V_0$ )

   b) Edges of the $G_0^*$ are not oriented, the set of edges of the $G_0^*$ includes only those edges of the $G_0$, which contains access rights Take or Grant ( $E_0^* = E_0 \setminus E_0^\alpha$ where $E_0^\alpha$ is the set of edges of the $G_0$ not containing access rights Take or Grant ).

   c) All edges of the $G_0^*$ has the same weight.

2) Applicate Dijkstra's algorithm to graph $G_0^*$. The shortest path between two vertices in the graph $G_0^*$ that found by the Dijkstra's algorithm will be the tg-path in a graph $G_0$.

The complexity of Dijkstra's algorithm depending on number of vertices is estimated as $O(N^2)$, where N is the number of vertices [4].

**Statement 1**. It is possible to build $G_0^*$ up to $O(N^2)$ operations.

**Proof**. Set of vertices of the graph $G_0^*$ is the same as set of vertices of the graph $G_0$, so $G_0^*$ differs from $G_0$ only the set of edges. Thus, to construct a graph $G_0^*$, we need to consider all the edges of the graph $G_0$. The maximum possible number of edges in the graph is $N(N-1)/2$, where N - the number of vertices. That is to construct a graph two need to do no more than $N(N-1)/2$ operations.

For an arbitrary protection graph it is necessary to find the island, bridges, and initially and terminally spans of the bridges.

To search the islands, we use Floyd's algorithm [3] as follows:

1) According to the original protection graph $G_0$ construct a graph $G_0^{'}$ as follows:

   a) Set of vertices of the graph $G_0^{'}$ is the same as set of vertices of the graph $G_0$ ( $V_0^{'} \equiv V_0$ ).
   
   b) Edges of the $G_0^{'}$ are not oriented, the set of edges of the $G_0^{'}$ includes only those edges of the $G_0$, which has subjects as start and final and contains access rights Take or Grant ( $E_0^{'} = E_0 \setminus E_0^{s-o} \cup E_0^{\alpha}$ where $E_0^{s-o}$ is the set of edges of the $G_0$ for which start and final are not both subjects and $E_0^{\alpha}$ is the set of edges of the $G_0$ not containing access rights Take or Grant ).
   
   c) All edges of the $G_0^{'}$ has the same weight.

2) Applicate Floyd's algorithm to graph $G_0^{'}$.

Floyd's algorithm finds all the shortest paths between every pair of vertices in the $G_0^{'}$. The shortest paths in the $G_0^{'}$ will be a tg-paths running through subjects in the $G_0$, and the subjects united by the tg-paths will be the islands in the $G_0$.

The complexity of Floyd's algorithm is estimated as $O(N^3)$ [4].

**Statement 2.** It is possible to build $G_0^{'}$ up to $O(N^2)$ operations.

**Proof.** The proof is analogous to Statement 1.

To search for bridges, as well as their initially and terminally spans, there are developed appropriate algorithms. The description of these algorithms is given in [5]. There two types of these algorithms are possible, one based on the depth-first search and second based on breadth-first search. The complexity of the algorithms does not exceed $O(N^4)$ ie the algorithms are polynomial.

## 4. Conclusions

Knowing the criteria of computer system state safety and knowing how to check these criteria, we can analyze a computer system for the presence of unauthorized access. The analysis can be performed automatically at the time not to exceed $O(N^4)$, where $N$ is the number of subjects and objects in a computer system.

Security analysis is relevant for all types of computer systems including operating systems, computer networks, database management systems and mobile systems. With the possibility of automatic analysis of security, we can create a more reliable systems with a lower costs.